\documentclass[12pt,a4paper]{aipproc}

\usepackage{graphicx}
\layoutstyle{6x9}

\begin{document}

\author{J.P.Hague}{address={Department of Physics, Loughborough University, Loughborough, LE11 3TU}}

\title{Extending the theory of phonon-mediated superconductivity in quasi-2D}


\begin{abstract}
I present results from an extended Migdal--Eliashberg theory of
electron-phonon interactions and superconductivity. The history of the
electron-phonon problem is introduced, and then study of the
intermediate parameter regime is justified from the energy scales in
the cuprate superconductors. The Holstein model is detailed, and
limiting cases are examined to demonstrate the need for an extended
theory of superconductivity. Results of the extended approximation are
shown, including spectral functions and phase diagrams. These are
discussed with reference to Hohenberg's theorem, the
Bardeen--Cooper--Schrieffer theory and Coulomb repulsion. {\bf [Published in: Lectures on the physics of highly correlated electron systems X, p255-264, AIP Conference Proceedings vol. 846 (2006)]}
\end{abstract}

\maketitle

\section{Introduction}

Over the past half-century, the study of the role of electron-phonon
interactions in condensed matter physics has been an active and
controversial field. Initially of interest from the point of view of
thermal properties, early models of the interactions between lattice
vibrations and electrons included the continuum Fr\"{o}hlich model
\cite{frohlich}. Interest in electron-phonon interactions increased
dramatically when in 1957, Bardeen, Cooper and Schrieffer (BCS)
published their famous theory of superconductivity
\cite{bcs}, which directly implicated phonons as the microscopic
mechanism for the low temperature absence of resistivity in a variety of
metals. Until the discovery of the cuprate superconductors by Bednorz
and M\"{u}ller in 1986 \cite{bednorz}, the BCS picture was found to
account well for all superconducting materials - a remarkable success
for a simple mean-field theory which is only applicable at weak
coupling!

Soon after the realisation that phonons were responsible for
superconductivity, Eliashberg extended the theoretical description
beyond the absolute weak coupling theory with the famous Eliashberg
equations \cite{eliashberg1960a}. In doing this, he built on the
earlier work of Migdal, who argued that a simple resummation of a
certain class of Feynmann diagrams should be sufficient to describe
the limit of low phonon frequency \cite{migdal1958a}. Eliashberg's
theory can be argued to be one of the first applications of the
dynamical mean-field theory (DMFT) \cite{metzner1989a}, since (in its original sense) it ignores
spatial fluctuations (momentum dependence) in the self-energy, while
keeping frequency dependent (dynamical) effects.

The purpose of this paper is to describe an extension to the
theory of superconductivity from electron-phonon interactions. The
approach goes beyond the Eliashberg theory by introducing the effects
of spatial fluctuations and higher order terms in the perturbation
theory. The aim is to develop a theory which can be used for systems
with stronger coupling, larger phonon frequencies and reduced
dimensionality. I begin by motivating the need for a more
sophisticated theory from the experimental viewpoint. I also discuss
limiting cases of the Holstein model, and how the large phonon
frequency limit of that model implies that the conventional theories
of superconductivity are incomplete. I then introduce the
approximations needed to develop a more sophisticated theory. Finally
I present some results from the new approximation, and discuss
them in relation to Cuprate superconductors, and also with regard
to conventional theories and the exact Hohenberg theorem
\cite{hohenberg}.

\section{Motivation}

When the high-temperature cuprate superconductors were discovered in
1986 \cite{bednorz}, the possibility that phonons could be attributed
to the microscopic mechanism was quickly discounted by many people. In
part, this was due to the absence of an isotope effect at optimal
doping, and also an assumption that phonon-mediated superconductivity
could not occur above 30K. The mechanism for
high-$T_C$ superconductors remains highly controversial, and many
different hypotheses are suggested (some examples are spin
fluctuations \cite{anderson} and exotic phonon mechanisms such as
bipolarons \cite{alexandrov}). An increasing body of evidence shows
that phonons as well as Coulomb repulsion have an effect on the
physics of the cuprate materials. I shall give a brief review of the
current experimental situation in this section, and argue that (1)
Electron-phonon interactions need to be treated on an equal footing to
Coulomb repulsion if the Cuprates are to be understood, and (2) In
order to treat the phonons in the Cuprates, extensions to the current
theories of electron-phonon interactions and phonon-mediated
superconductivity are required.

\begin{figure} 
\includegraphics[width=14cm]{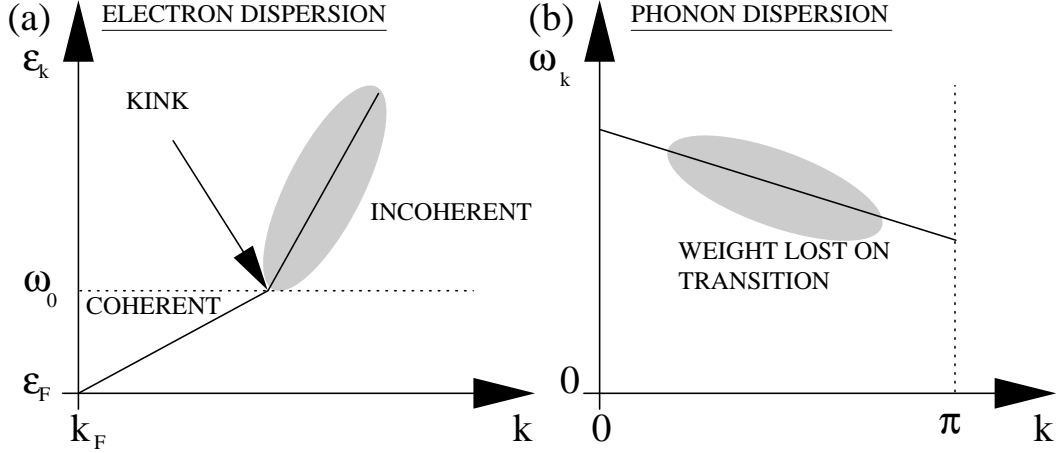}
\caption{Schematics showing the effect of electron-phonon interactions
  on the electron and phonon dispersions in the cuprates. Both panels
  describe measurements along the [11] direction. Panel (a) shows a
  schematic representation of the electronic dispersion measured by
  Angle-Resolved Photo-Emmission Spectroscopy (ARPES)
  \cite{lanzara2001a}. At energies close to the Fermi-surface, there
  are coherent excitations with a long lifetime. As
  $\epsilon_k=|\omega_0-\epsilon_F|$ is approached, the gradient of
  the dispersion changes and a kink is introduced. The phonon is of
  the transverse optic variety, and its frequency ($\omega_0$) is
  $\sim 75$meV. The ratio of the gradients above and below the kink is
  related to the coupling constant \cite{lanzara2001a}. Panel (b)
  shows a schematic representation of some neutron scattering results
  measuring the phonon dispersion
  \cite{mcqueeny1999a,chung2003a}. Above the transition temperature,
  this looks like the solid line, but as the system moves from the
  normal to the superconducting state, the spectral weight in the
  shaded area vanishes. This indicates that the superconducting state
  affects the phonons, and is further evidence for strong
  electron-phonon coupling.}
\label{fig:dispersion}
\end{figure}

There are several experiments demonstrating strong electron-phonon
coupling in the cuprates. The most compelling is the existence of a
strong isotope effect on exchanging O$^{16}$ for O$^{18}$
\cite{zhao1997a}. There are also some more recent experiments which
demonstrate the effects of electron-phonon interactions in a
transparent manner. Figure \ref{fig:dispersion} shows schematic
representations of electron and phonon dispersions in the
cuprates. Panel (a) details the main features of the electronic
dispersion measured by Angle-Resolved Photo-Emmission Spectroscopy
(ARPES) in the [11] direction \cite{lanzara2001a}. At energies close
to the Fermi-surface, there are coherent excitations with a long
lifetime. As $\epsilon_k=|\omega_0-\epsilon_F|$ is approached, the
gradient of the dispersion changes at a sharp kink. The phonon is of
the transverse optic variety, and its frequency ($\omega_0$) is of the
order of 100meV. It suffices here to mention that this is very
large. The ratio of the gradients above and below the kink is related
to the dimensionless coupling constant
($\lambda= g^2/t\omega_{0}$), and it is found that $\lambda$
can take values of up to 2 \cite{lanzara2001a}. Panel (b) shows a
schematic representation of some neutron scattering results measuring
the phonon dispersion \cite{mcqueeny1999a,chung2003a}. Above the
transition temperature, this looks like the solid line, but as the
system moves from normal to superconducting state, the spectral weight
in the circled area vanishes. This indicates that the
superconductivity (bound pairs of electrons) affects the phonons, and
is additional evidence for a strong electron-phonon coupling.

\begin{figure}
\includegraphics[width=10cm]{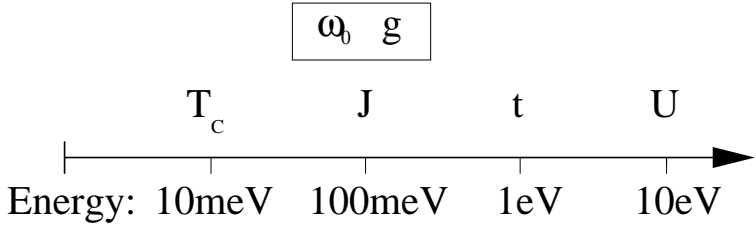}
\caption{Schematic showing the energy scales in the cuprates. The
  largest energy by far is the Coulomb repulsion (or Hubbard $U$) of
  order 10eV. The intersite hopping integral $t$, is $\sim$1eV. Using
  a simple 2nd order perturbation theory, an effective exchange
  interaction is generated, with $J=t^2/U$ of the order of
  100meV. This $J$ is then used to argue for the spin-fluctuation
  theory of high $T_C$. However, the energies of the phonons are also
  approximately 100meV and the dimensionful coupling $g$ has around
  the same value. Thus with 3 similar energy scales, it is important
  that the contributions from both spin-fluctuations and phonon
  mechanisms are treated on equal footing.}
\label{fig:energies}
\end{figure}

A frequent misconception about the cuprates is that electron-phonon terms
in the Hamiltonian can be neglected on the basis that they are
small. To demonstrate that this is not the case,
figure \ref{fig:energies} shows approximate energy scales in the
cuprates. The largest energy by far is the Coulomb repulsion (or
Hubbard $U$) which weighs in at some 10eV. Next is the intersite
hopping integral $t$, which is of the order of 1eV. Using a simple 2nd
order perturbation theory at strong coupling, an effective exchange
interaction is generated \cite{gebhard}, with $J=t^2/U$ of the order
of 100meV. This $J$ often used to argue for a spin-fluctuation theory
of high-$T_C$ superconductivity that neglects phonons. The problem
with this viewpoint is immediately clear if one reviews the
experimental data. First, the energies of the phonons are also
approximately 100meV, so they cannot be treated as a small energy
scale. Second, a dimensionless coupling constant of order unity
implies dimensionfull coupling $g$ with similar magnitude. Thus with
three very close energy scales, it is important that the contributions
from both phonon and Coulomb mechanisms are treated on equal footing
in a theory for the cuprates. Unfortunately, as I discuss in the next
section, current theories of electron-phonon interactions are not
capable of handling the large phonon energies and coupling constants
in the cuprates. The remainder of this paper focuses on how the theory
can be extended to describe this regime.

\section{Model and limits}

A generic model of electron-phonon interactions includes the motion of
the electrons $H_{\mathrm{el}}$, the motion of the ions (or phonons)
$H_{\mathrm{ph}}$ and the interaction between the electrons and the
phonons (which may be absorbed or emitted) which is denoted $H_{\mathrm{el-ph}}$. In
this way, $H=H_{\mathrm{el}}+H_{\mathrm{el-ph}}+H_{\mathrm{ph}}$ is
the total Hamiltonian.
\begin{equation}
H_{\mathrm{el}}=\sum_{\mathbf{k}}\epsilon_{\mathbf{k}}c^{\dagger}_{\mathbf{k}}c_{\mathbf{k}}\approx-\sum_{<ij>\sigma}t c^{\dagger}_{i\sigma}c_{j\sigma}
\end{equation}
\begin{equation}
H_{\mathrm{el-ph}}=-\sum_{\mathbf{q},\mathbf{k}}\bar{g}_{\mathbf{q}}c^{\dagger}_{\mathbf{k}-\mathbf{q}}c_{\mathbf{k}}(b^{\dagger}_{\mathbf{q}}+b_{-\mathbf{q}})\approx-\sum_{i\sigma} n_{i\sigma} gr_i
\end{equation}
\begin{equation}
H_{\mathrm{ph}}=\sum_{\mathbf{k}}\omega_{\mathbf{k}}\left(b^{\dagger}_{\mathbf{k}}b_{\mathbf{k}}+\frac{1}{2}\right)\approx\sum_i\left( \frac{M\omega_{0}^2r_i^2}{2}+\frac{p_i^2}{2M}\right)
\end{equation}
The first term in the Hamiltonian is the general form for free
electrons, i.e. the total energy is the sum of the kinetic energies of
all occupied states. In a special case, which is known as the Holstein
Hamiltonian, the electrons in a tight binding model may hop between
nearest-neighbour sites only, and
$\epsilon_{\mathbf{k}}=-2t\sum_{i=1}^{D}\cos(k_{i})$, where $t$ is the
overlap integral. In the generic form of the electron-phonon
interaction, an electron may be scattered by absorbing a phonon with
momentum $-\mathbf{q}$ or emitting a phonon with momentum
$\mathbf{q}$. An additional approximation uses a momentum independent
electron-phonon coupling, $g$, and in that case the Fourier transform
shows that the second term connects the local ion displacement, \(
r_{i} \) to the local electron density. Finally, the free phonon term
may be simplified by using the Einstein approximation
$\omega_{\mathbf{k}}\approx\omega_0$ and Fourier transforming, the
bare phonon Hamiltonian is shown to be a series of independent simple
harmonic oscillators at each site index. The creation of electrons and
phonons is represented by $c^{\dagger}$ and $b^{\dagger}$
respectively, \( p_{i} \) is the ion momentum and \( M \) the ion
mass. By choosing $t=0.25$, a bandwidth of $W=2$ is chosen. A small
interplanar hopping of $t_{\perp}=0.01$ is included to remove the
logarithmic singularity in the 2D density of states at $\epsilon=0$.

\begin{figure}
\includegraphics[width=8cm]{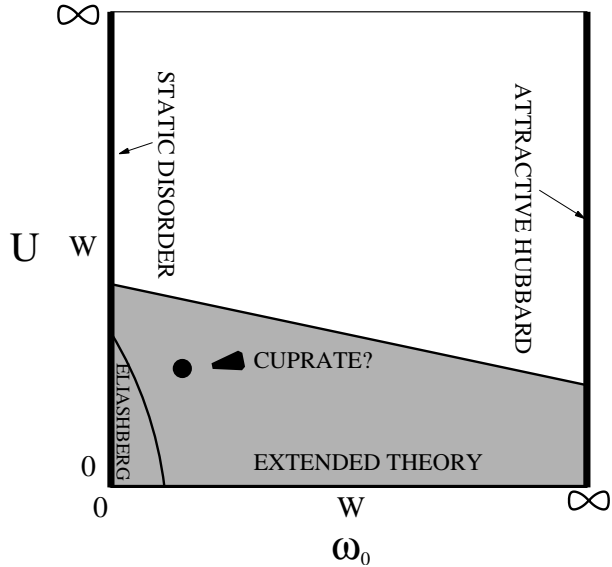}
\caption{Parameter space of the Holstein model. For very large phonon
  frequency, the effective interaction is instantaneous, and a
  Lang--Firsov transformation results in an attractive Hubbard
  model. Alternatively, taking the limit of very small phonon
  frequency, a fast moving electron cannot `see' the phonons move, and
  the problem maps to a static disorder problem (similar to the
  Falikov--Kimball model \cite{millis1996a}). This makes the phonon
  problem extremely hard, and little is known about the middle of the
  parameter space. The range of the Eliashberg theory is shown in the
  bottom left corner. The expected position of the cuprates is shown
  as the single diamond. The expected validity of an extended theory
  including all 2nd order Feynman diagrams is also shown.}
\label{fig:parameterspace}
\end{figure}

Figure \ref{fig:parameterspace} shows the parameter space of the
Holstein model. For very large phonon frequency, the effective
interaction is instantaneous, and a Lang--Firsov transformation
\cite{langfirsov} results in an attractive Hubbard model (which is one
of the standard models for correlated electron systems)
\cite{hubbard1963a}. Alternatively, taking the limit of very small
phonon frequency, a fast moving electron cannot `see' the nuclei move
in the time it takes to traverse many sites, so the problem maps to a
static disorder problem (which is essentially uncorrelated). One may
therefore think of the phonon frequency as possessing the ability to
``tune'' the effect of correlations, and one therefore obtains a
second motivation for the study of electron-phonon systems of trying
to understand electronic correlations \cite{haguenda}. The correlation
tuning makes the phonon problem extremely hard, and little is known
about the intermediate regime of the parameter space. The range of the
Eliashberg theory is shown in the bottom left corner. Contrary to
Migdal's assumption, the theory cannot extend beyond intermediate
coupling since renormalisation of the effective mass reduces
$\epsilon_F$ invalidating the condition (Migdal's theorem)
$\omega_0\ll\epsilon_F$ \cite{hague2001a,alexandrov}. The approximate
position of the phonon parameters in the cuprates is shown as the
single diamond. It is essential to correct the theory for weak to
intermediate coupling at larger phonon frequencies. The extension is
clear by looking at the large phonon frequency limit. The Hubbard
limit requires that all 2nd order processes in $U$ are included in the
self-energy, or the incorrect weak coupling limit is found. An
extended theory including all 2nd order Feynman diagrams is required
to understand the weak coupling limit, from small to large phonon
frequency.

\section{Extending the Eliashberg theory}

\begin{figure}
\includegraphics[width=8cm]{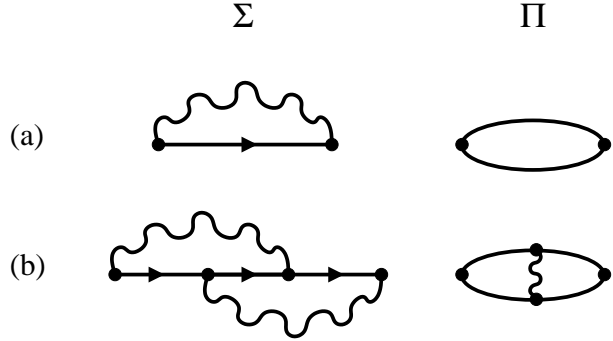}
\caption{Series of Feynman diagrams used in the current approximation. $\Sigma$ is the electron and $\Pi$ the phonon self-energy. Series (a) is the Migdal-Eliashberg approximation and (b) the vertex corrected series.}
\label{fig:feynman}
\end{figure}

Extending the Eliashberg theory involves inserting the lowest order
vertex corrections into the electron and phonon self energies. In the
Eliashberg theory, emitted phonons are reabsorbed in a
last-out-first-in order. Vertex corrections essentially allow this
order to be changed once. Such contributions are shown diagrammatically
in figure \ref{fig:feynman}. All the diagrams must be included in the
calculation, or electron number would not be conserved. Momentum
dependence is included in the approximation, which is essential in
low-dimensions. The inclusion of vertex corrections leads to double
2-fold integration over the Brillouin zone in combination with a
double sum over matsubara frequencies, which is time consuming for the
numerics. In order to reduce the number of points in
$\mathbf{k}$-space while maintaining the thermodynamic limit, the
dynamical cluster approximation is applied
\cite{hettler1998a}. Additionally, superconducting states can be
considered by using the Nambu formalism. The full details of the
implementation of the extended approximation can be found in
references \cite{hague2003a} and \cite{hague2005d}.

\begin{figure}
\includegraphics[height=14cm,angle=270]{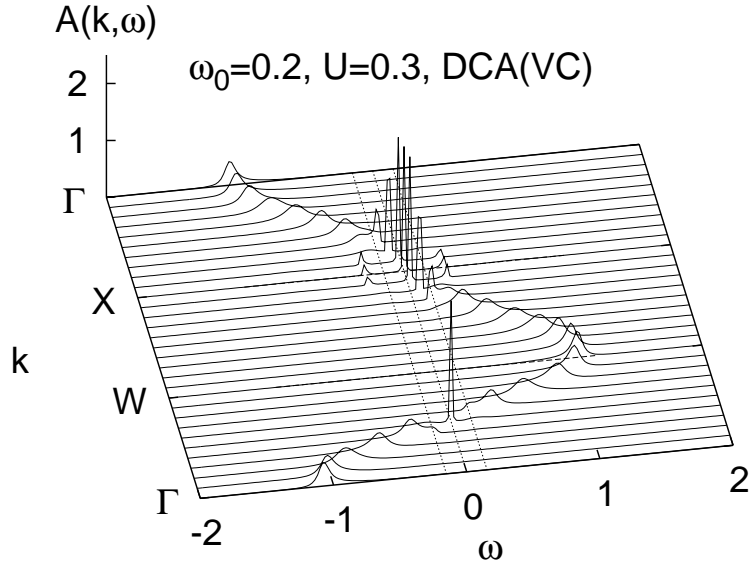}
\caption{Spectral function of the Holstein model in the extended
Migdal--Eliashberg theory. The results are qualitatively similar to
ARPES measurements of the cuprates. In particular the change between
incoherent and coherent particles occurs at the phonon frequency,
associated with a kink in the $[11]$ direction. \copyright Institute of
physics publishing 2003 \cite{hague2003a}.}
\label{fig:spectral}
\end{figure}

Using a maximum entropy technique, it is possible to compute the
spectral function from the Matsubara axis Green function. Figure
\ref{fig:spectral} shows the spectral function of the Holstein model
calculated using the extended Migdal--Eliashberg theory. The results
are qualitatively similar to ARPES measurements of the cuprates. In
particular the change between incoherent and coherent particles occurs
at the phonon frequency (shown as the dashed line), associated with a
kink in the $[11]$ direction. It is noted here that the effect of the
phonon self-energy is a softening of the phonon mode. In the standard
ME theory in 2D, the mode at the $(\pi,\pi)$ point is completely
softened, leading to a fatal instability of the theory. However, the
vertex corrections act against this softening, and relieve the
instability. In such a way, it is clear that a vertex corrected
Eliashberg theory is essential for the study of quasi-2D materials
\cite{hague2003a}.

\begin{figure}
\includegraphics[height=10cm,angle=270]{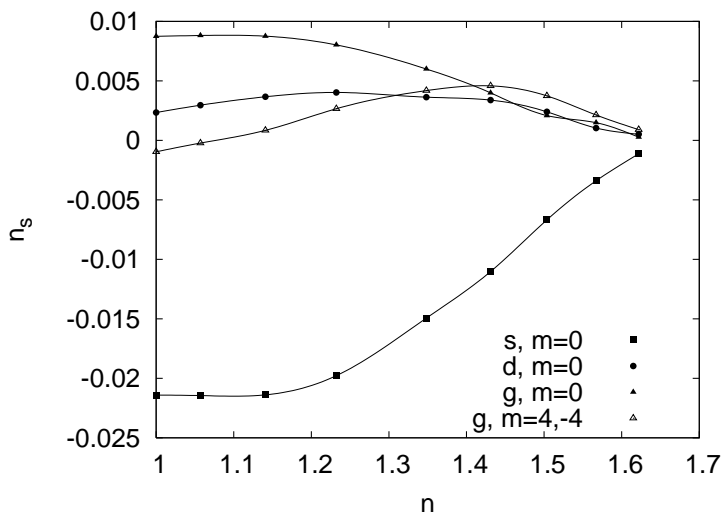}
\caption{Decomposition of the order parameter into spherical
harmonics. A cluster size, $N_C=64$ is used, with $U=0.6$ and
$\omega_0=0.4$. Note how higher order harmonics develop as the filling
is increased. In particular, the $g$ harmonics can be almost as strong as the $s$ harmonics at $n=1.45$. \copyright Institute of physics publishing 2005 \cite{hague2005d}.}
\label{fig:decomposition}
\end{figure}

One can also compute properties in the superconducting state. One such
property is the momentum-dependent pairing density,
$n_s(\mathbf{k})=T\sum_n F(i\omega_n,\mathbf{k})$, where
$F(i\omega_n,\mathbf{k})$ is the anomalous Green function associated
with the pairing of electrons with momentum $\mathbf{k}$ and
$-\mathbf{k}$. It is possible to transform the momentum dependent
order parameter to determine the magnitude of individual spherical
harmonics. Figure \ref{fig:decomposition} shows such a
decomposition. A cluster size of $N_C=64$ is used, with $U=0.6$ and
$\omega_0=0.4$. Note how higher order harmonics develop as the filling
is increased. In particular, it can be seen that no single harmonic
(such as the $s$-wave symmetry) is sufficient to describe the order
parameter. Some of the higher order terms come about due to increased
pairing at momentum $\mathbf{k}=(\pi/2,\pi/2)$, in particular, pairs
with angular momentum.

\begin{figure}
\includegraphics[width=9cm]{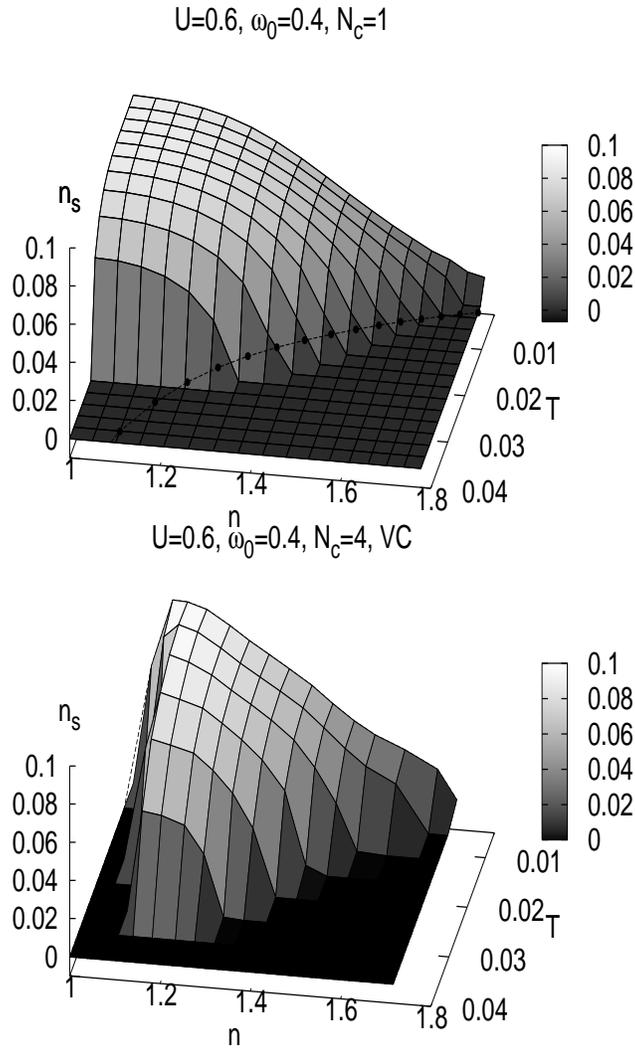}
\caption{Phase diagrams of the Holstein model. $U=0.6$ and
$\omega_0=0.4$. The top diagram shows the result from the Eliashberg
approximation (dynamical mean-field theory $N_C=1$). Also shown is the
BCS result (line with points). On the bottom the results from the
current approximation with $N_C=4$ are shown. The superconducting
order is suppressed close to half filling in the vertex corrected
theory. \copyright Institute of physics publishing 2005 \cite{hague2005d}.}
\label{fig:phasediagram}
\end{figure}

Finally, by varying the temperature and chemical potential, the phase
diagram can be computed. Figure \ref{fig:phasediagram} shows phase
diagrams of the Holstein model for the different
approximations. $U=0.6$ and $\omega_0=0.4$.The top diagram shows the
result from the Eliashberg approximation (dynamical mean-field theory
$N_C=1$). On the bottom the results from the current approximation with
$N_C=4$ are shown. The superconducting order is suppressed close to
half filling. Assuming a form for the density of states in 2D (with
small interplane hopping) of
${\mathcal{D}}(\epsilon)=(1-t\log((\epsilon^2+t_{\perp}^2)/16t^2))/t\pi^2$
(for $|\epsilon|<4t$) \cite{liliana}, which matches the full density
of states with reasonable accuracy. From this the BCS result may be
calculated using the expression
\begin{equation}
T_C(n)=2\omega_0\exp(-1/|U|{\mathcal{D}}(\mu(n)))/\pi, 
\label{eqn:tc}
\end{equation}
with the chemical potential taken from the self-consistent solution
for a given $n$. This result also drops off monotonically. Results in
the dilute limit are in good agreement with the BCS result (line with
points). Close to half-filling, the DMFT result is significantly
smaller than the BCS result (which predicts $T_C(n=1)>0.07$). The
difference in results between the two mean-field theories at
half-filling is due to the self-consistency in the DMFT. When vertex
corrections and spatial fluctuations are included, the dilute limit is
relatively unchanged. However at half-filling, there is a huge drop in
the transition temperature. The suppression at half-filling is a
manifestation of Hohenberg's theorem, which implies that there may be
no superconducting order in 2D. Here I have computed for quasi-2D, so
it is interesting that in real materials with low dimensional
character the maximum in superconductivity is shifted away from
half-filling.

\section{Concluding remarks}

I end the paper with a warning for constructing theories of
high-temperature superconductivity using electron-phonon interactions
alone, while neglecting the Coulomb repulsion. If one takes the phase
diagrams from the previous section, and assigns similar energy scales
to those in the cuprates, it is possible to obtain a temperature in
Kelvins for the maximum in the phase diagram at $n=1.2$. This comes
out as around 172K - one could say approximately the $T_C$ in the
cuprates.

So why isn't this the solution for the cuprates? Cuprates are very
tightly bound materials, which is why the ``Fermi energy'' is low, and
the ratio $\omega_0/\epsilon_F$ is large enough to justify extending
Eliashberg theory. The problem is that a small Fermi energy also means
the the Hubbard $U$ is a comparatively large quantity. On a simple
mean-field level, one can include the Coulomb repulsion in the theory
of superconductivity. For example, the Eliashberg equations can be
extended to include an effective electron-electron interaction
(otherwise known as the Coulomb pseudopotential $\mu_C$). The effect
of this is to modify $\lambda\rightarrow\lambda-\mu_C$. Substitution
into equation \ref{eqn:tc} means that the transition temperature is
considerably reduced, or that superconductivity of the BCS type is
completely destroyed. Any phonon-based mechanism for the cuprates must
address this point and be compatible with the electron-electron
interaction.  Alternatively (and this is a warning against the other
extreme) on the basis of the similarity of energy scales, any
spin-fluctuation mechanism (which is essentially Coulombic) must also
treat the phonons (or at least be compatible with them) to be
plausible.

\section{Acknowledgments}

I sincerely thank the organising committee of the course for their
generous financial support. Aspects of this research were carried out
under the MPIPKS guest scientist program, and as a visitor at the
University of Leicester. I thank A.S.Alexandrov, J.L.Beeby, E.M.L.Chung,
N.d'Ambrumenil, J.K.Freericks, M.Jarrell, P.E.Kornilovitch, J.H.Samson
and M.Yethiraj for stimulating discussions, both about this work and
the problems of electron-phonon interactions and superconductivity in
general. I acknowledge support at Loughborough University under EPSRC
grant no. EP/C518365/1.

\bibliographystyle{unsrt}
\bibliography{salerno_proceedings}

\end{document}